\documentstyle[12pt,psfig,aasms4]{article}

\received{November 00 2001}
\accepted{April 25 2002}

\begin{document}

\title{Chandra observation of Abell 1689:
New determination of mass distribution and 
comparison to lensing measurements}

\author{Sui-Jian Xue and Xiang-Ping Wu}

\affil{National Astronomical Observatories,
Chinese Academy of Sciences, Beijing 100012, China}

\begin{abstract}
We present a new estimate of the projected X-ray mass of
Abell 1689 observed with Chandra, 
in an attempt to clarify the issue of whether or not 
there exists a discrepancy between X-ray and gravitational
lensing mass estimates claimed in previous investigations based on
{\it Einstein}, {\it ROSAT} and {\it ASCA} observations. 
A particular attention is paid to examining 
if there is an offset between X-ray centroid and
central dominant cD galaxy, which may be an indicator of 
the presence of local dynamical activities of intracluster gas 
in the central core and therefore,  explain
the discrepancy between X-ray and lensing mass estimates, if any.
The unprecedentedly high spatial resolution achieved  
by Chandra allows us to precisely localize the X-ray centroid
of  Abell 1689, which appears to coincide perfectly with 
the central cD galaxy. 
This fact, along with the symmetry and regularity of the X-ray surface 
brightness and temperature distributions, 
suggests that Abell 1689 is a fully-relaxed cluster. 
We thus employ hydrostatic equilibrium hypothesis
to determine the projected mass profile of Abell 1689, and compare
it with the results obtained by different lensing techniques
available in the literature.  Our analysis confirms the existance
of the discrepancy of a factor of $\sim2$ 
between X-ray and lensing mass estimates  
in the central region of $r\approx0.2$ Mpc, although the two methods 
yield essentially consistent result on large radii.
If the perfect coincidence between the X-ray center and
the cD galaxy of Abell 1689 detected by Chandra observation  
is not a projection effect, the central mass discrepancy 
between X-ray and lensing measurements may pose a challenge to
our conventional understanding of dynamical evolution of 
the intracluster gas in the central regions of clusters. 
\end{abstract}


\section{Introduction}

Clusters of galaxies serve as an ideal cosmological laboratory for test of 
theories of formation and evolution of structures, and for determinations
of fundamental cosmological parameters such as the matter density parameter
$\Omega_M$, the baryon fraction $f_b$, the normalization parameters
$\sigma_8$ and the shape parameter $\Gamma$. All these cosmological 
applications depend critically on how accurately the mass distributions
of clusters can be measured. Current techniques include the optical
measurements of distribution and velocity dispersion of cluster 
galaxies combined with the Jeans equation, 
the X-ray observations of intracluster gas 
(density and temperature profiles) coupled with the hydrostatic equilibrium
hypothesis, and the gravitational lensing effects of distant galaxies behind
clusters. Other methods such as the measurements of hot intracluster gas 
through the Sunyaev-Zeldovich effect and the 
numerical simulations of formation 
and evolution of clusters also act as a complementary manner to the 
above direct approaches. 
A comparison of different mass estimates  
provides a simple way to examine the reliability and accuracy of
different methods, which also allows one to diagnose dynamical evolution 
of different matter components (galaxies, gas and dark matter) of clusters.
In such an exercise, the mass distribution revealed by gravitational lensing,
which is independent of the matter content and dynamical state of clusters,
may be used as a reference point for judging other approaches.
Previous studies have essentially arrived at the following conclusions:
There is good agreement between the weak gravitational lensing, 
X-ray and optical determined cluster masses on large scales characterized
by X-ray core radii, while in the central regions of some lensing 
clusters the X-ray method is likely to systematically underestimate 
cluster masses by a factor of $\sim2$ 
(Wu 1994; Miralda-Escud\'e \& Babul 1995; 
Wu \& Fang 1997; Allen 1998; Wu et al. 1998; Wu 2000 and references therein).
Because this discrepancy occurs in the very central cores of clusters,
it may arise from our oversimplification of dynamical properties of 
the intracluster gas on small scales.
Indeed, the discrepancy becomes significant only for non-cooling flow clusters
(Allen 1998; Wu 2000), in which the apparent offsets 
between X-ray and lensing centers have been detected (Allen 1998;
Katayama, Hayashida \& Hashimotodani 2002). Therefore, detailed knowledge
of the gas distribution in the inner regions of clusters will be 
crucial for clarifying the issue.

The unprecedentedly high spatial resolution of the Chandra X-ray 
Observatory has permitted detailed studies of intracluster gas in
the innermost regions of clusters. So far, central X-ray masses have been 
explicitly derived for several lensing clusters observed with Chandra. 
For at least three clusters (Abell 1835, MS 1358  and Abell 2390), 
the newly determined X-ray masses are found to be in good agreement with
the gravitational lensing measurements (Allen, Ettori \& Fabian 2001;
Schmidt, Allen \& Fabian 2001; Arabadjis, Bautz \& Garmire 2001),
while a significant discrepancy between the X-ray and strong lensing
mass estimates is also detected in Abell 2218 (Machacek et al. 2001).
For the latter, the X-ray centroid shows an offset of 
$\sim 22^{\prime\prime}$ from the dominant cD galaxy. This reinforces
the argument that the mass discrepancy may be attributed to the 
inappropriate use of hydrostatic equilibrium to the intracluster
gas in the central regions where substructures and their bulk motion
may govern the dynamical process of intracluster gas even if clusters
become fully relaxed. 
Indeed, the existence of subsonic motion of central gas
has been conformed recently by Chandra observation in  cluster RXJ1720
which has a redshift (0.164) similar to those of lensing clusters
(Mazzotta et al. 2001). Such motion may explain the offsets between 
X-ray and lensing centers. In a word, it seems that the consistency
or discrepancy between the two mass estimates can be simply related to
the question of whether or not the X-ray centers show a displacement from 
the central cD galaxies in lensing clusters.

In this paper, we present a Chandra observation of the lensing 
cluster Abell 1689 at redshift $z=0.181$,  
in an attempt to reexamine the consistency/discrepancy
between X-ray and lensing mass estimates and the possible offset
between X-ray and optical centers.
The mass distribution of Abell 1689 has been reconstructed by several 
independent techniques of gravitational lensing: arcs and arclets
(Tyson \& Fischer 1995), the number count dilution of lensed red
galaxies (Taylor et al. 1998), the distortion of background
galaxy luminosity function (Dye et al. 2001) and the weak lensing
of background galaxies (Clowe \& Schneider 2001). 
This cluster was observed with {\sl EINSTEIN}, {\sl ROSAT}, 
{\sl GINGA}, and {\sl ASCA}. 
The X-ray luminosity in the 0.1 - 2.4 band is 	
$(2.74\pm3.32)\times10^{45}$ erg s$^{-1}$ (Ebeling et al. 1996)
and the X-ray temperature is $T=9.02^{+0.40}_{-0.30}$ keV 
(Mushotzky \& Scharf  1997). The X-ray determined
mass within the giant arc position ($\sim51^{\prime\prime}$), 
under the assumption of hydrostatic 
equilibrium, is found to be lower by a factor of $\sim2$ than  
that derived from strong gravitational lensing (Miralda-Escud\'e \& 
Babul 1995; Wu \& Fang 1997; Allen 1998; Wu 2000). In particular,
based on the {\sl ROSAT HRI} observation,
Allen (1998) attributed this discrepancy to the offset of 
$13^{\prime\prime}.3$ between the lensing and X-ray centers.
The offset value is within $1-2\sigma$ position uncertainty 
considering that {\sl ROSAT HRI} provides a $\sim5$ arcsec
(full width at half-maximum) imaging resolution. 
With the new Chandra observation of Abell 1689, 
which is one magnitude improved in spatial resolution with 
respect to {\sl ROSAT HRI}, 
we are now able to reexamine the reported mass discrepancy and
its origin. Throughout this paper , we
assume $H_0=50$ km s$^{-1}$ Mpc$^{-1}$ and $\Omega_0=1$.

\section{Chandra observation and analysis}

\subsection{Data preparation}
Abell 1689 was observed twice for separate 10 ksec by 
the Chandra X-ray Observatory on April 15, 2000 
and January 7, 2001, respectively. Both observations were
performed by using mainly the on board Advanced CCD Imaging 
Spectrometer in  2$\times$2 imaging array (ACIS-I) mode. 
In the present work, the dataset is 
retrieved from the Chandra archive for the first 10 ksec observation. 
The second observation, unfortunately, suffered from some serious
aspect problem and the dataset has not yet been made available 
for public. 

In order to achieve a better modeling of instrument gain and
quantum efficiency, we process the Level 1 CXC data products using 
Chandra Interactive Analysis of Observations
package CIAO-2.2, with complement of the updated 
calibration database CALDB-2.9.
Following the standard methods (Chandra Science Threads for CIAO-2.2\footnote
{available at http://asc.harvard.edu/ciao/documents\_threads.html.}), 
we remove the Pixel and PHA Randomization effects, and ``streak'' 
events from the original event file. 

Periods of background flaring are removed using the CIAO task {\bf lc\_clean}
for ACIS chips I0-3 one by one. This finally yields an effective 
exposure time of $\sim7182$ s after all known corrections for 
Abell 1689 are taken into account.

While the telescope pointing placed bulk of the cluster emission
on the ACIS chip I3, a certain cluster emission also
extends over other three adjacent chips. This may cause some 
problems in finding source-free background regions 
for the spatial and spectral analysis. Thus, a suitable blank-sky 
background dataset, acisi\_C\_i0123\_bg\_evt\_230301.fits, compiled by 
Markevitch (2001)
 has been chosen and tailored for the present
observation. We apply the same cleaning algorithm to the data as was applied
to the background set. Identical spatial and energy filters are applied
to source and background data so that the background normalization is set
by the ratio of their exposure times. Moreover, 
we have created a spectral weighted exposure map (see section 2.2) 
for the entire ACIS chip array, 
using the aspect histogram files which contain information about 
aspect motion during the observation for each chip, and instrument maps 
which are essentially the product of detector quantum efficiency and 
mirror effective area projected onto the detector surface.

Figure 1 displays a true color image of Abell 1689, produced by
a mosaic of three X-ray bands,
red: 0.3--1.5\,keV, green: 1.5--3\,keV and blue: 3--10\,keV.
Images in each of the three bands, after being normalized by the 
exposure map, have been binned by four (such that an image pixel subtends 
1.96 arcsec$^2$ on the sky) and adaptively smoothed 
using CIAO task {\it csmooth}
at 3$\sigma$ significant levels with a maximum smoothing scale of 5 pixels.
It appears that with Chandra's high spatial resolution a large number
of X-ray point sources have been resolved within the field around the galaxy 
cluster, $\sim16'\times16'$.  
 
In order to check the Chandra astrometry, we use the CIAO tool
{\bf wavdetect}, with a significance parameter value of $10^{-7}$,
to search for sources in the broad band image (0.3-10 keV) of the cluster
field. A total of 29 sources are detected (see Figure 1). 
The source list is then compared with  
the optical USNO-A2.0 catalog and the radio FIRST catalog . Using
a correlation radius of $1.5''$, we find three matches within
a $12'$ radius of the ACIS-I center. The cumulative probability of a
single random association between this data is 0.016. The probability
that all the three matches occurred by chance is only 3.8$\times10^{-6}$. 
Therefore, we conclude that the astrometry of this Chandra observation 
is not worse than $1''.5$ without further adjustments.

\subsection{Spectral analysis}
First we wish to work out the mean temperature and metallicity of the 
cluster. For this purpose, an overall spectrum within a large 
circular region of radius $3.7'$ 
(corresponding to a physical radius of 0.87 Mpc) has been extracted
from the cleaned events after excluding several point sources. 
The resulted spectrum is further grouped so that each energy bin contains at 
least 20 counts. Because for energies below 0.7 keV  uncertainties in 
the current available ACIS-I response functions are relatively large,
our spectrum analysis is restricted to the 0.7-9 keV energy band. 
We fit the spectrum using the thermal emission model MEKAL
with a fixed absorption at the Galactic value,
1.8$\times10^{20}$ $\rm cm^{-2}$. We find that this model (hereafter Model 1) 
does not give the best description of the data reflected by  
$\chi^{2}$/d.o.f$=344/311$. 
The best fit of the data needs an additional absorption to the MEKAL
model (hereafter Model 2), which improves the fit of Model 1 significantly
at $>99.9$\% confidence level with $\chi^{2}$/d.o.f$=289/310$.
The observed and fitted spectra of Abell 1689 are plotted in Figure 2, 
and the best fit parameters of both models are listed in Table 1. 
It turns out that although the resulting X-ray temperature depends 
sensitively upon whether or not there is an extra absorption component,
the best fit value of metallicity remains almost unaffected. This last
point can be seen in the confidence plot (Figure 3) of
metallicity versus temperature. 
In summary, the average temperature and metal abundance for Abell 1689
are $kT= 8.2-10.0$ keV (or $11.0-14.7$ keV) and $Z=0.20-0.49$ of solar,
respectively.
The bulk emission flux in the 0.5--10 keV band is  $2.7\pm1.0\times10^{-11}$ 
$\rm ergs\ cm^{-2}\ s^{-1}$, corresponding to an X-ray luminosity of 
$L_x=(4.12\pm1.53)\times10^{45}$ erg s$^{-1}$ in the  0.5--10 keV band
or a bolometric X-ray luminosity of 
$L_x=(6.65\pm2.55)\times10^{45}$ erg s$^{-1}$.  This is in good 
agreement with the expected value from the statistical $L_x$-$T$ relation 
for clusters (Wu, Xue \& Fang 1999) when the flux falling out the 
detection aperture is included using the $\beta$ model given below.

\subsection{Temperature structure}
Now we investigate the temperature structure of the cluster. We extract 
spectra for a series of annuli centered at the centroid of cluster emission
(see section 2.4 for spatial analysis). 
All these spectra are properly grouped, which allows us to perform
the minimum\ $\chi^{2}$ analysis. Here,
both Model 1 and Model 2 are used. Alternatively, 
in order to set a robust constraint on temperature value, we freeze the 
metallicity parameter in the fitting process. The results are summarized in 
Table 2 and the azimuthally averaged temperature profile is shown
in Figure 4.

\subsection{Surface brightness}
In order to take the possible asymmetric emission distribution into account,
we first model the bulk X-ray emission of the cluster with a 
two-dimensional $\beta$ model in the form
\begin{equation}
S_x(x,y)=S(r)=\frac{S_{x0}(x_0,y_0)}{(1+(r/r_c)^2)^{\alpha}},
\end{equation}
where 
\begin{equation}
r(x,y)=\frac{\sqrt{x'^2(1-\epsilon)^2+y'^2}}{1-\epsilon},
\end{equation}
and
\begin{eqnarray}
x'&=&(x-x_0)\cos\theta+(y-y_0)\sin\theta,\\
y'&=&(y-y_0)\cos\theta+(x-x_0)\sin\theta.
\end{eqnarray}
The model assumes that cluster emission has a projected elliptical 
brightness surface with ellipticity $\epsilon$ and position angle $\theta$.
The surface brightness is measured in photon counts/arcsec$^2$/s  
with respect to the X-ray centroid ($x_0,y_0$), and is characterized by
the core radius $r_c$ and  a power-law index $\alpha$.
We apply this model to the exposure corrected 
image of the cluster in 0.3-10 keV, and also include
the blank field background.
The best fit results are summarized in Table 3. It appears that
the core radius is in the range of $r_c=16''.6-18''.6$, the index $\alpha$ is 
1.0--1.2, the normalization is $S_{x0}=(6.7-7.6)\times10^{-6}$ photons\ 
$\rm cm^{-2}$ s$^{-1}$, and the X-ray emission centroid is located at
$\alpha=13^h11^m29^s.45 (\pm3.45'')$, 
$\delta=-01^{\circ}20'28''.06 (\pm1.35'')$.
Overall, the emission pattern is rather symmetric about the
centroid with a very small value of ellipticity, $\epsilon=0.08$. 

We have performed a detailed check of the HST observations of Abell 1689,
which clearly shows a central dominant cD galaxy, with the position 
coinciding nicely with the X-ray centroid within $\sim1.5''$ 
uncertainties. We have then checked the FIRST observation in 
the radio band, which demonstrates a central emission peaked 
at ($13^h11^m30^s.002,-01^{\circ}20'28''.28$). This is about
$8''$ from the X-ray emission centroid. A visual examination shows
that the radio  centroid is centered on
one of the companion galaxies around the central cD. In Figure 5, we 
plot the X-ray emission contours overlaid on an HST/WFPC2  
image of Abell 1689.

Since the X-ray emission of the cluster is fairly symmetric, we 
first employ the conventional single $\beta$ model to fit the 
radial brightness profile of the cluster. However, the fitting is not
acceptable,  reflected by $\chi^2/{\rm d.o.f}=225.4/49$.
We then adopt a double $\beta$ model
\begin{equation}
S_x(r)=S_{x0}^1\left(1+\frac{r^2}{r_{c1}^2}\right)^{-3\beta_1+0.5} +
       S_{x0}^2\left(1+\frac{r^2}{r_{c2}^2}\right)^{-3\beta_2+0.5},
\end{equation}
which provides a significantly
reduced $\chi^2$ fit to the data with $\chi^2/{\rm d.o.f}=50.6/46$. 
The best fit parameters are summarized in Table 4, and 
the observed and fitted surface brightness profiles are 
demonstrated in Figure 6.

\section{Mass determinations} 

Morphology of the overall X-ray surface brightness distribution of A1689,
along with the nearly constant temperature across the cluster surface out to
$r\sim1$ Mpc, 
suggests that Abell 1689 is a dynamically-relaxed cluster. This argument
is further supported by the fact that the X-ray centroid coincides with 
the central dominant cD galaxy. Consequently, one may use
the hydrostatic equilibrium hypothesis to evaluate the dynamical
mass of the cluster. For an isothermal, double $\beta$ model,  
the total dynamical mass within radius $R$ can be obtained from
\begin{equation}
M(R)=-\frac{kTR^2}{G\mu m_p n(R)}
     \left[\frac{dn_1(R)}{dR}+\frac{dn_2(R)}{dR}\right],
\end{equation}
where $\mu=0.59$ is the average molecular weight, $n_i(R)$ (i=1,2) is
the particle number density for the $i$th phase gas and 
$n(R)=n_1(R)+n_2(R)$. By reverting the observed surface brightness 
distribution $S_x(r)$ [equation (5)], 
we can derive an analytic expression for $n_i(R)$ 
(Xue \& Wu 2000). Next, we infer the mass density profile from
$\rho(R)=(1/4\pi R^2)[M(R)/dR]$, and calculate the projected 
mass $m(r)$ within radius $r$ from the cluster center along the 
line-of-sight. The resulting mass distribution based on the average
temperature of $T=13.2^{+1.5}_{-1.2}$ from Model 1 is shown in
Figure 7. Note that utilization of Model 1 yields a maximum estimate of 
the X-ray cluster mass because the temperature given by Model 1 is 
significantly higher than that by Model 2.

On the other hand, 
the projected cluster mass within the position of giant arc $r_{\rm arc}$
can be estimated through
\begin{equation}
m(r_{\rm arc})=\pi r^2_{\rm arc}\Sigma_{\rm crit},
\end{equation}
where $\Sigma_{\rm crit}=(c^2/4\pi G)(D_{\rm s}/D_{\rm d}D_{\rm ds})$ 
is the critical surface mass density, $D_{\rm d}$, $D_{\rm s}$ and 
$D_{\rm ds}$ are the angular diameter distances to the cluster, 
to the background galaxy, and from the cluster to the galaxy, respectively. 
Unfortunately, the redshift of the giant arc at $51^{\prime\prime}$ 
from the cD galaxy of Abell 1689 remains unknown.  
Assuming $z_s=0.8$ and $z_s=2.0$
for the arc in the above equation yields
$m(r_{\rm arc})=3.76\times10^{14}M_{\odot}$ and
$m(r_{\rm arc})=3.18\times10^{14}M_{\odot}$, respectively. For comparison,
the X-ray cluster mass within $r_{\rm arc}$ derived from
the Chandra data reads 
$m(r_{\rm arc})=2.44^{+0.25}_{-0.20}\times10^{14}M_{\odot}$ and
$m(r_{\rm arc})=1.66^{+0.21}_{-0.10}\times10^{14}M_{\odot}$ for
Mode1 and Model 2, respectively. 
Namely, the ratios of strong lensing masses to X-ray masses given
by spectral Model (1, 2) within $r_{\rm arc}$ 
are ($1.54^{+0.14}_{-0.14}$, $2.27^{+0.15}_{-0.26}$) 
and ($1.30^{+0.12}_{-0.12}$, $1.92^{+0.11}_{-0.22}$) for $z_s=0.8$ 
and $z_s=2.0$, respectively. 
All the quoted errors are $90\%$ confidence limits. 
Recall that previous estimates based on the {\sl ROSAT} and {\sl ASCA} data 
($T\approx9.0$ keV) found a mass ratio of $\sim2$ (Allen 1998; Wu 2000),
which is in good agreement with our results based on  Model 2. 
In other words, the high resolution observation of Chandra
yields essentially the same result of X-ray cluster mass in the central
core of Abell 1698. Only for Model 1 can the mass discrepancy 
be reduced especially when the background source is assumed at large 
redshifts and the measurement uncertainties are included.

The projected mass distribution of Abell 1689 has also been obtained from
the study of the deficit of red galaxies 
behind the cluster by Taylor et al. (1998).
We illustrate their reconstructed mass profile in Figure 7, 
allowing the uncertainties to include $90\%$ confidence limits.
Another mass estimate is provided by Dye et al. (2001) from the comparison
of luminosity functions of background galaxies  
between the lensed field behind Abell 1689 and an undistorted control field. 
Their cumulative projected masses within three radii are shown in Figure 7.
Finally, Clowe \& Schneider (2001) fitted their weak lensing derived
cluster masses to a singular isothermal sphere for a set of assumed 
background galaxy redshifts, which are also 
demonstrated  in Figure 7.  
It appears that the overall projected cluster masses reconstructed by 
four different lensing methods (arcs, number counts, distorted 
luminosity function of galaxies and weak lensing analysis)
show a good agreement within measurement uncertainties.

The newly determined mass of Abell 1689 in the central region 
of $r\approx0.2$ Mpc by Chandra observation
is systematically lower than those derived from strong and weak lensing
techniques, although the two innermost data points given by
the distorted luminosity function of galaxies are still consistent with
the projected X-ray mass when the large measurement uncertainties 
are included. Nevertheless, at large radii $r>0.6$ Mpc the two mass 
estimates yield roughly the same result.

\section{Discussion and conclusions}

The high spatial resolution observation of Abell 1689 with Chandra 
has permitted an accurate localization of the X-ray centroid 
of the cluster, which shows a perfect coincidence with the central 
cD galaxy, suggesting that the cluster has become fully relaxed at
its redshift $z=0.181$. The central cooling time within
$r=10$ kpc is $t_{\rm cool}=2.54^{+0.20}_{-0.17}$ Gyr and 
the cooling radius is $r_{\rm cool}=150$ kpc.
The newly determined X-ray surface brightness and temperature
distributions, incorporated with the hydrostatic equilibrium 
hypothesis, have allowed us to re-evaluate  
the projected mass profile of the cluster and compare it with 
the results provided by different lensing techniques. 
It shows that the significant mass discrepancy
between X-ray and lensing methods reported in previous studies
still remains even if the new Chandra data are used. 

With the unprecedented sub-arcsec spatial resolution, as well
as the large effective detection areas in the broad energy range
of 0.3-10 keV,
the new Chandra observation of Abell 1689 disagrees with  
previous findings based on {\sl ROSAT} observation 
that there exists an offset of $13^{\prime\prime}.3$ 
between the X-ray and optical centers (Allen 1998).
The latter was suggested to be 
the main cause for the discrepancy between X-ray and lensing
mass estimates. Indeed, the presence of such offset  
is a good indicator of local violent activities
of the intracluster gas in the central core so that hydrostatic
equilibrium may become inapplicable. An alternative is
that the offset is produced by the oscillation of the central
cD galaxy around the bottom of the cluster potential well
(Lazzati \& Chincarini 1998).
The perfect coincidence between the X-ray center and
the cD galaxy of Abell 1689 detected by Chandra observation, 
if it is not a projection effect, and the central mass discrepancy 
between X-ray and lensing measurements pose a challenge to
the above speculation.  If confirmed, this would have significant
impacts on our understanding of the distribution and evolution 
of intracluster gas in the central cores of clusters. 
Future observations of other lensing clusters with Chandra 
would clarify the issue.

\acknowledgments

We thank the referee for constructive suggestions.  
This work was supported by
the National Science Foundation of China, under Grant No. 19725311
and the Ministry of Science and Technology of China, under Grant
No. NKBRSF G19990754.

\clearpage


\begin{deluxetable}{ccccc}
\small
\tablecaption{Average Emission properties of Abell 1689 }
\tablehead{
\colhead{Model} & \colhead{$N_{\rm H}$} &\colhead{Temperature}  & 
\colhead{Metallicity} &\colhead{$\chi^{2}$/d.o.f} \\
\\
& \colhead{($10^{20}{\rm cm}^{-2}$)} & \colhead{(keV)} & \colhead{(Z$_{\odot}$)} 
& }
\startdata
Model 1 & -  & 13.2$^{+1.5}_{-1.2}$ & 0.35$^{+0.14}_{-0.15}$ & 344/311 \nl 
Model 2\tablenotemark{a} &6.7$^{+1.5}_{-1.5}$ & 9.0$^{+1.0}_{-0.8}$ & 0.31$^{+0.10}_{-0.09}$ & 289/310 \nl 
\enddata
\tablenotetext{a} {Observed flux in 0.5-10 keV, 2.7$^{+0.1}_{-0.1}
\times10^{-11}$ $\rm ergs\ cm^{-2}\ s^{-1}$.}
\tablecomments{All the fittings have a fixed absorption at the Galactic
value. All the errors quoted are at the 90\% confidence level 
(i.e. $\Delta \chi^{2}=2.7$).}
\end{deluxetable}

\clearpage

\begin{deluxetable}{ccll}
\small
\tablecaption{Temperature structures of Abell 1689 }
\tablehead{
\colhead{Annuli($\arcsec$)} & \colhead{Net counts} & \colhead{T$_1$\tablenotemark{a}\ (keV)} &\colhead{T$_2$
\tablenotemark{a}\ (keV)}  
 }
\startdata
 0-10   & 1197  & 14.5(10.4-20.8)  & ~9.7(7.1-13.8) \nl
10-20   & 2074  & ~8.7(~7.3-10.5)    & ~7.8(6.7-9.3) \nl
20-30   & 1818  & ~9.1(~7.6-11.5)    & ~7.6(6.5-9.2) \nl
30-40   & 1219  & 17.6(12.5-24.3)  & 14.3(9.2-17.1) \nl
40-50   & 1388  & 12.4(~9.2-16.4)   & ~8.8(6.9-11.1) \nl
50-70   & 1804  & 11.8(~9.2-15.7)   & ~8.5(7.0-10.8) \nl
70-80   & 921   & 14.3(10.0-21.8)  & 11.3(8.0-16.7) \nl
80-115  & 1850  & 13.5(10.4-18.2) & ~9.8(8.0-12.6) \nl
115-155 & 1521  & 14.8(11.1-20.6) & 12.9(9.6-17.6) \nl
\enddata
\tablenotetext{a} {Temperature values are from Model 1 and Model 2, 
respectively. The intervals in parentheses represent 90\% confidence limits.}
\end{deluxetable}

\clearpage

\begin{deluxetable}{cccccc}
\scriptsize
\tablecaption{2-dimensional Modeling of the surface brightness profile 
of Abell 1689 }
\tablehead{
\colhead{$r_c$} & \colhead{$\alpha$} & \colhead{$S_{x0}$}
& \colhead{$x_0,y_0$} & \colhead{$\epsilon$} & \colhead{$\theta$} \\ 
($\arcsec$) &  & (photons\ cm$^{-2}$ s$^{-1}$) & ($\alpha, \delta$) & & (degree) \nl
 }
\startdata
 $17.6\pm1.0$ & $1.1\pm0.1$  & $(7.1\pm0.5)\times10^{-6}$ & 
$13^h11^m29^s.45(\pm3.45''), -01^{\circ}20'28''.06 (\pm1.35'')$  & 0.08 &
1.7 \nl
\enddata
\end{deluxetable}

\begin{deluxetable}{cccccc}
\small
\tablecaption{Double $\beta$ Modeling of the surface brightness profile
of Abell 1689 }
\tablehead{
\colhead{$r_{c1}$} & \colhead{$\beta_1$} & \colhead{$S^1_{x0}$}
& \colhead{$r_{c2}$} & \colhead{$\beta_2$} & \colhead{$S^2_{x0}$} \\
($\arcsec$) &  & (photons\ cm$^{-2}$ s$^{-1}$) & 
($\arcsec$) &  & (photons\ cm$^{-2}$ s$^{-1}$)  
\nl
 }
\startdata
 $20.6\pm0.4$ & $0.72\pm0.01$  & $(4.1\pm0.1)\times10^{-7}$ &
 $90.0\pm1.0$  & $0.87\pm0.01$ & $(4.5\pm0.1)\times10^{-8}$ \nl
\enddata
\end{deluxetable}

\clearpage

\begin{figure}
\epsscale{1.0}
\plotone{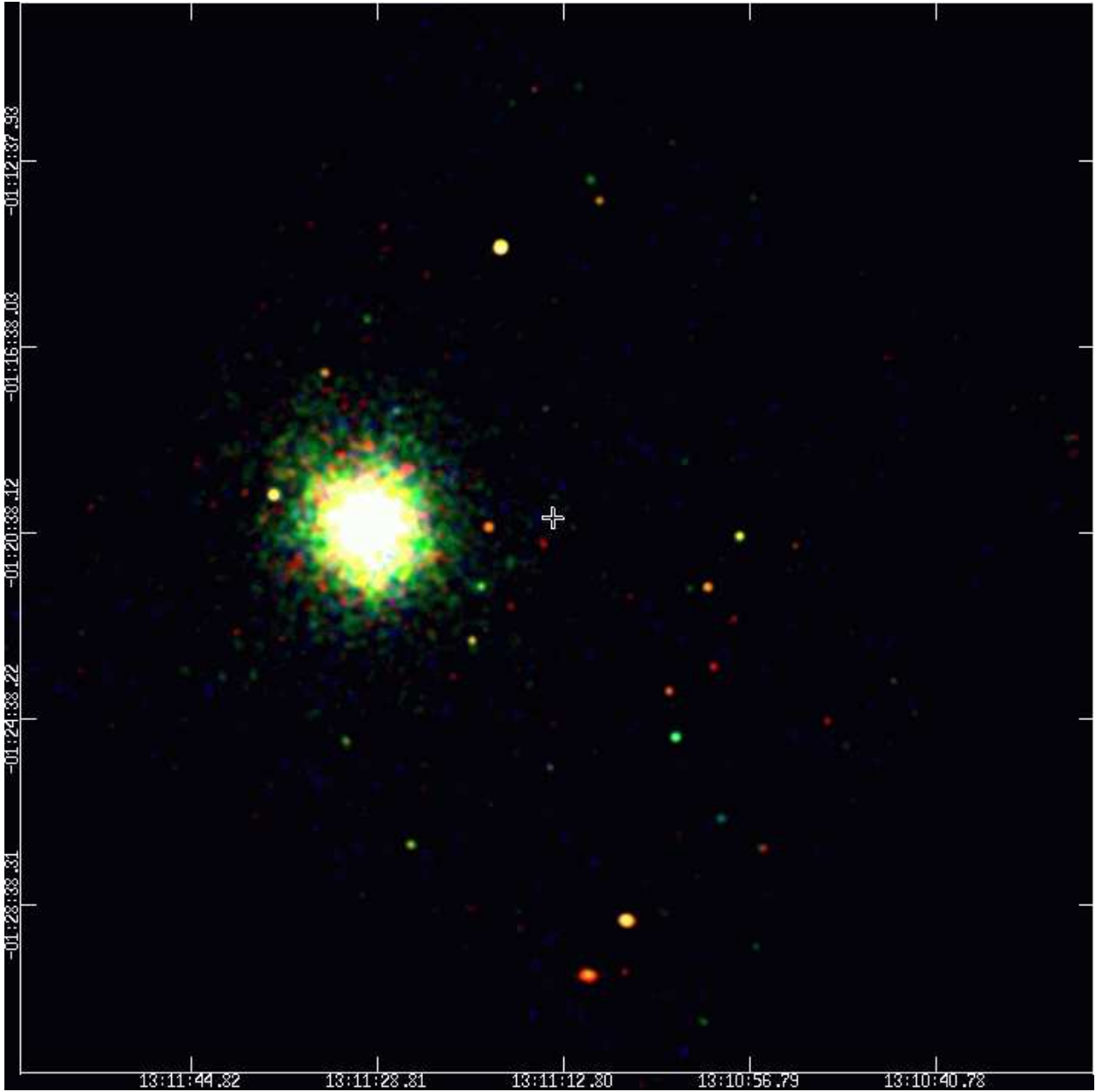}
\caption{True color image of Abell 1689,
produced by a mosaic of three smoothed X-ray bands:
red: 0.3--1.5\,keV; green: 1.5--3\,keV; and blue: 3--10\,keV.
Numerous X-ray point sources have been resolved within
the field around the galaxy cluster, $\sim16'\times16'$.
\label{fig1}}
\end{figure}

\begin{figure}
\epsscale{1.0}
\plotone{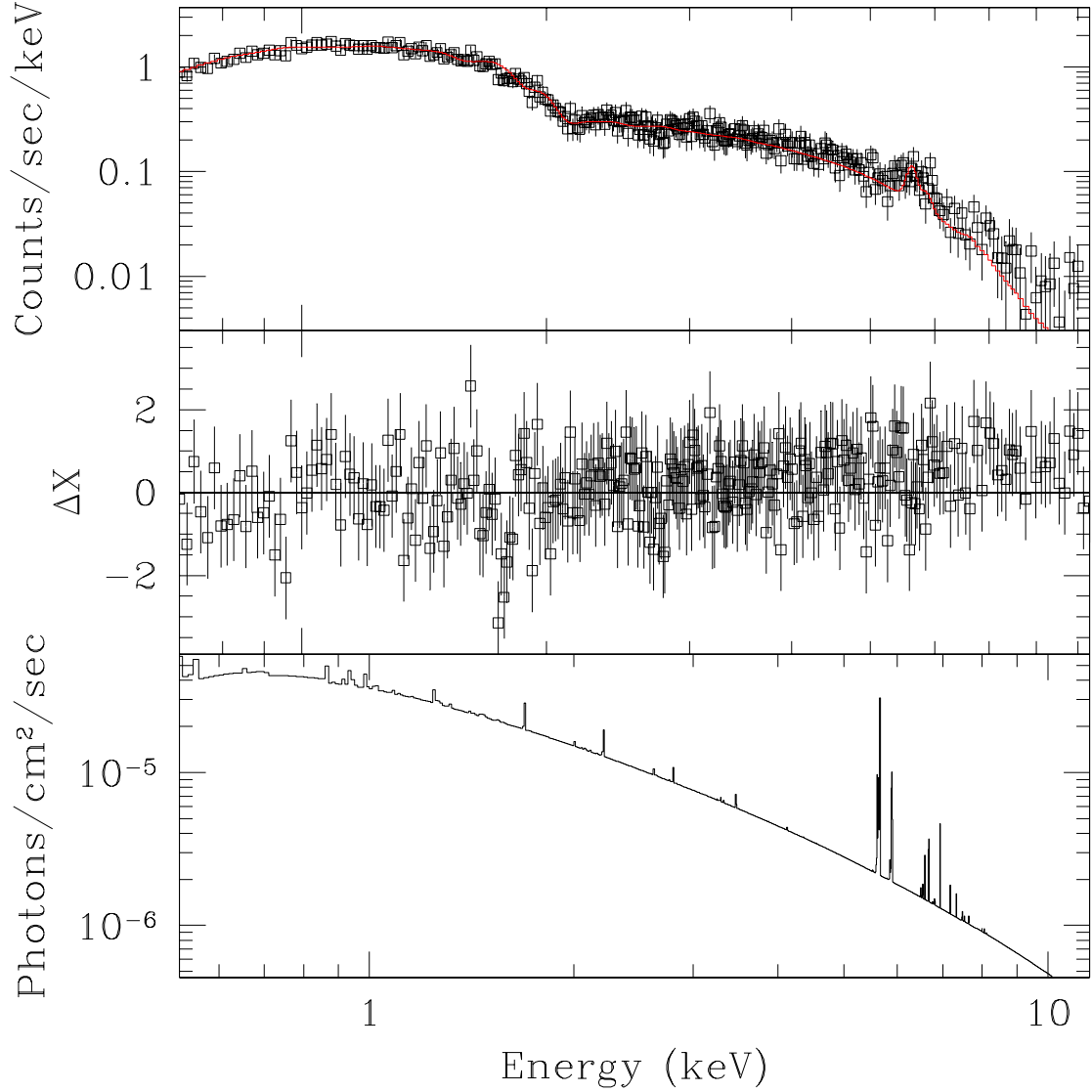}
\caption{Observed X-ray spectrum of Abell 1689 (upper panel)
fitted by the thermal emission model MEKAL with
an excess absorption (Model 2, bottom panel) convolved with Chandra
instrumental responses. The fitting residuals are shown in 
the middle panel. 
\label{fig2}}
\end{figure}

\begin{figure}
\epsscale{0.8}
\plotone{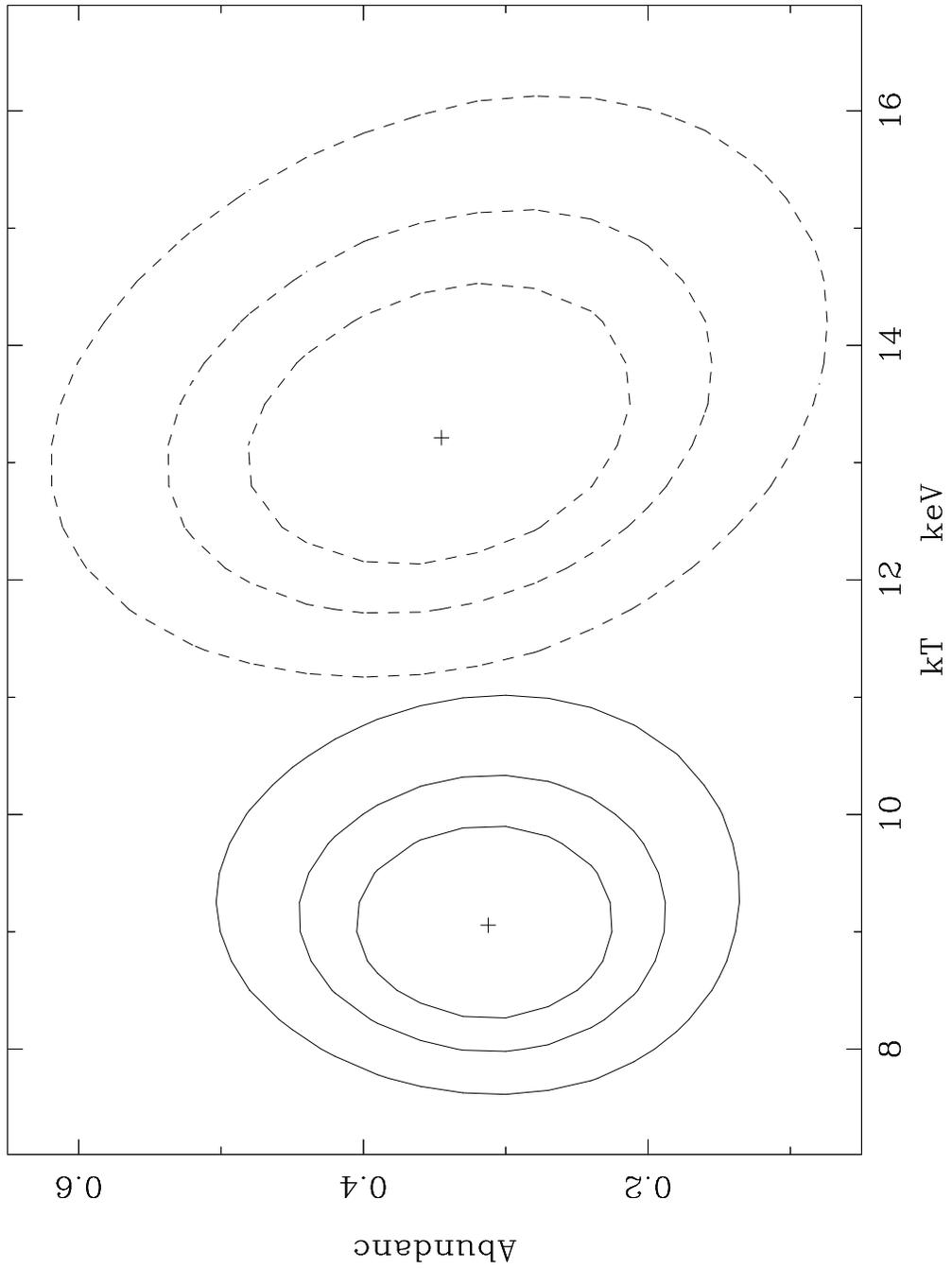}
\caption{Confidence plot (68\%, 90\%, and 99\%) of metallicity versus 
temperature for bulk thermal emissions of Abell 1689. 
Dashed-line and solid-line contours are for Model 1 and Model 2,
respectively.  
\label{fig3}}
\end{figure}

\begin{figure}
\epsscale{1.0}
\plotone{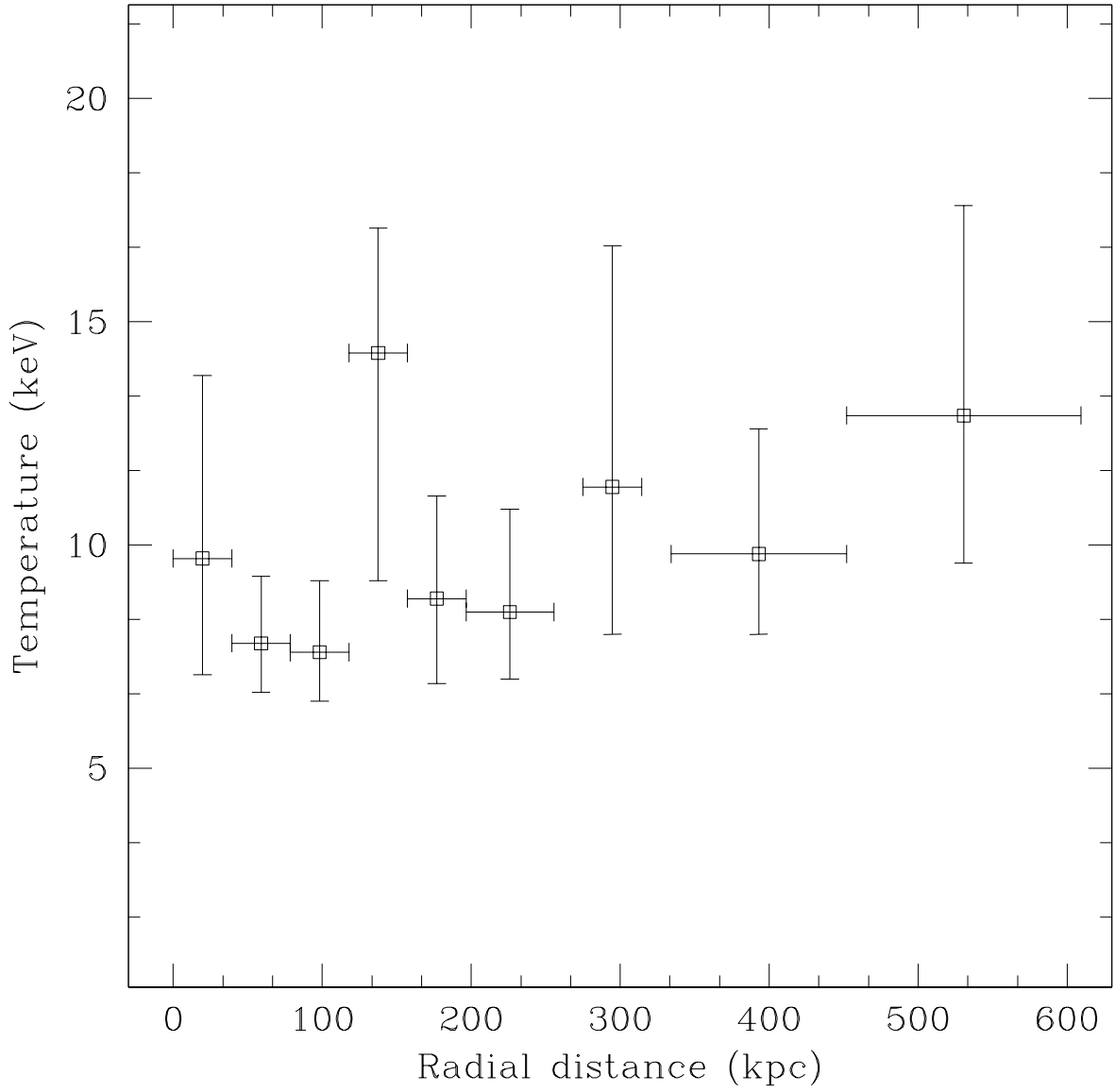}
\caption{Temperature profile of Abell 1689 constructed from Model 2.
\label{fig4}}
\end{figure}

\begin{figure}
\epsscale{1.0}
\plotone{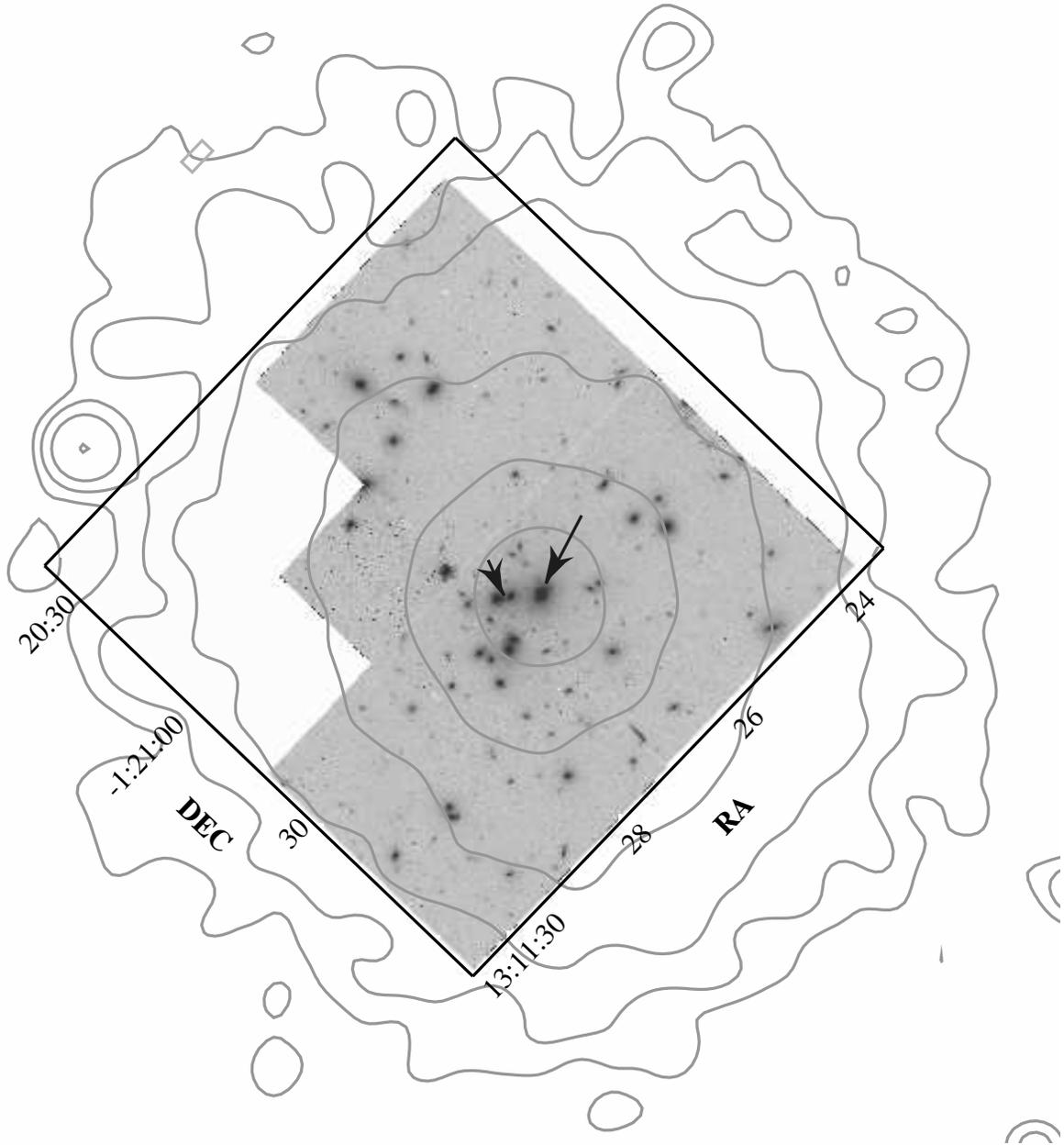}
\caption{X-ray emission contours overlaid on an HST/WFPC2 image of the
central region of Abell 1689. Up is the North and left is to the East.
Note that the central cD galaxy of the cluster, as indicated by the big arrow,
coincides perfectly with the X-ray emission centroid.
The small arrow points the position of the radio (FIRST) peak.
\label{fig5}}
\end{figure}

\begin{figure}
\epsscale{1.0}
\plotone{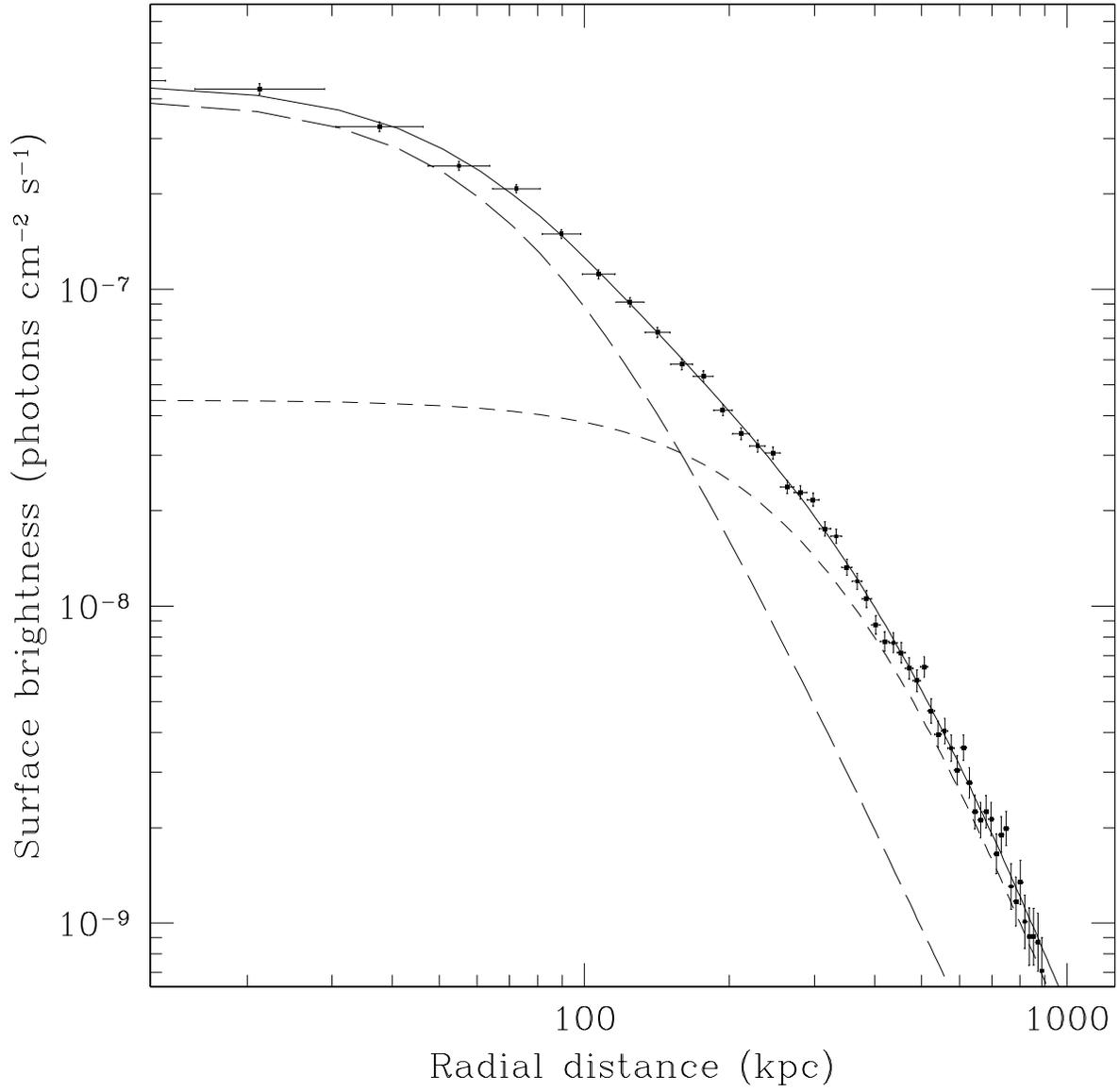}
\caption{Observed radial profile of X-ray surface brightness,
together with the double $\beta$ model fitting.
\label{fig6}}
\end{figure}

\begin{figure}
\epsscale{0.8}
\plotone{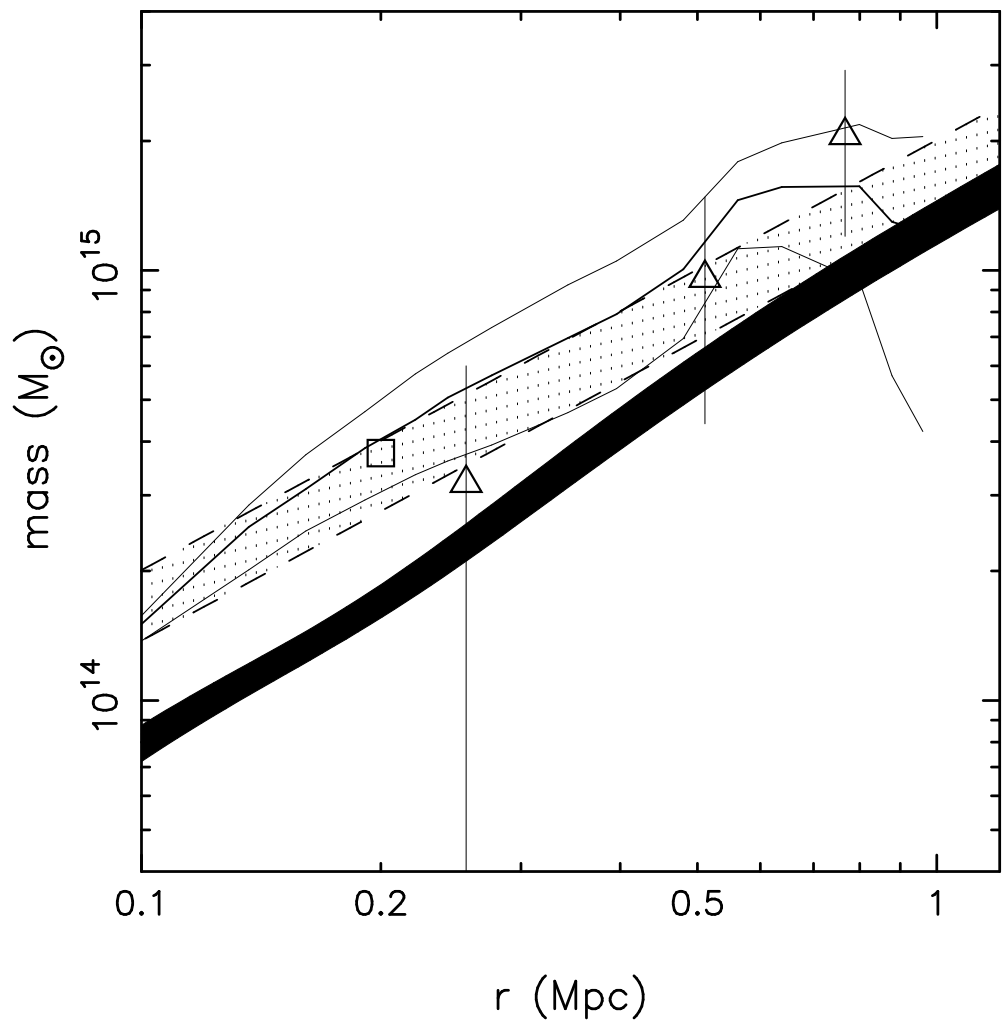}
\caption{Projected mass distributions of Abell 1689 derived from
the Chandra observation (filled region with $90\%$ confidence limits),
the strong gravitational lensing
(square), the distortion of background galaxy luminosity
function (triangles), the deficit of number counts of red galaxies
(solid lines) and the best fit singular isothermal sphere to
the weak lensing data for background galaxy redshifts from
$z=0.5$ to $3.0$ (dashed lines).
\label{fig7}}
\end{figure}

\end{document}